\documentclass[reprint, showpacs, amsmath, amssymb, aps]{revtex4-1}
\usepackage{mathrsfs}
\usepackage{amsfonts}
\usepackage{amsmath,amsfonts,amssymb,mathrsfs,bm}
\usepackage{verbatim,psfrag,times,CJK}
\usepackage{graphicx,graphics,color,epsfig}
\usepackage{float}
\usepackage{flafter}
\usepackage{subeqnarray}
\usepackage{cases}
\usepackage{amsmath, upgreek, amssymb, graphicx, subfigure, paralist, dsfont}
\usepackage[colorlinks, linkcolor=blue, citecolor=blue, urlcolor=blue, breaklinks=true]{hyperref}

\definecolor{red}{rgb}{1,0,0}
\newcommand{\ket}[1]{\left\vert{#1}\right\rangle}
\newcommand{\bra}[1]{\left\langle{#1}\right\vert}

\begin{document}

\title{Mirror and cavity formations by chains of collectively radiating atoms}

\author{Qurrat-ul-Ain \surname{Gulfam}}
\email{qgulfam@jazanu.edu.sa} 
\affiliation{Department of Physics, Faculty of Science, Jazan University, P.O. Box 114, Gizan 45142, Saudi Arabia}
\author{Zbigniew \surname{Ficek}}
\email{zficek@kacst.edu.sa} 
\affiliation{The National Centre for Applied Physics, KACST, P.O. Box 6086, Riyadh 11442, Saudi Arabia}
\affiliation{Quantum Optics and Engineering Division, Institute of Physics, University of Zielona G\'ora, Szafrana 4a, Zielona G\'ora 65-516, Poland}

\date{\today}

\begin{abstract}
We search for mirror and cavity-like features of a linear chain of atoms in which one of the atoms is specially chosen as a probe atom that is initially prepared in its excited state or is continuously driven by a laser field. Short chains are considered, composed of only three and five atoms. The analysis demonstrate the importance of the inter atomic dipole-dipole interaction which may lead to a collective ordering of the emission along some specific directions. We examine the conditions under which the radiative modes available for the emission are only those contained inside a cone centered about the inter atomic axis. Particular interest is in achieving the one-way emission along the inter atomic axis, either into left (backward) or right (forward) direction, which is referred to as a mirror-like behavior of the atomic chain. 
A direction dependent quantity called the directivity function, which determines  how effective the system is in concentrating the radiation into a given direction, is introduced. We show that the function depends crucially on the distance between the atoms and find that there is a threshold for the inter atomic distances above which a strongly directional emission can be achieved. The one-sided emission as a manifestation of the mirror-like behavior and a highly focused emission along the inter atomic axis as a characteristic of a single-mode cavity are demonstrated to occur in the stationary field. Below the threshold the directivity function is spherically symmetric. However, we find that the population can be trapped in one of the atoms, and sometimes in all atoms indicating that at these small distances the system decays to a state for which there are no radiative modes available for emission. 
\end{abstract}


\maketitle

\section{Introduction}

Advancement in the current technology of trapping and controlling single atoms cooled down to ultra low temperatures has opened new research directions in quantum optics and quantum communication.  
Spatial configurations of linear atomic chains or two-dimensional atomic lattices have been engineered and have been widely applied in various experimental setups~\cite{coldgas,ryd1,ryd2}. Recently, the subject of utilizing super-cold atoms as highly reflecting mirrors has gained much attention. In particular, it has been demonstrated that a collection of cold atoms trapped near the surface of a one-dimensional waveguide can form a nearly perfect mirror for the radiation incident on the atoms~\cite{QST,cj12,Chang}. 
 The waveguide represents a photonic channel which enhances the electromagnetic field to which the atoms are coupled thereby leading to a strong collective behavior of the atoms. As a consequence, a large part of the incident light is directed, reflected back, to the medium from which it originated. This striking mirror property of atoms is in contrary to the usual observation where atoms absorb/scatter all or most of the incident energy. 
 
Another kind of systems that can exhibit mirror properties or equivalently a highly directive radiative properties are atoms chirally coupled to a waveguide~\cite{pr15,vr16}. Chirality in atom-waveguide coupling is an effect associated with a broken symmetry of emission of photons from the atoms into the right and left propagating modes of the waveguide. As a result, the emitted photons are channeled into one of the two directions of the waveguide. It has been shown that the chiral property of the emission can enhance entanglement between two distant atoms~\cite{gg15}.
 Directive radiative properties have also been demonstrated for a single atom trapped at front of a distant dielectric mirror~\cite{hs11}. It has been demonstrated both theoretically and experimentally that the atom can behave as an optical mirror effectively forming, together with the dielectric mirror, a Fabry-P\'erot cavity. Related studies have shown that An atom mirror cannot only serve as a single mirror for a one-dimensional cavity, but also could be arranged to behave like a high-finesse cavity~\cite{supercavity}. 

In the course of previous work on directional emission the underlying atoms independently couple to a one-dimensional field of a waveguide, or nano-cavity or a nano-fibre. 
Although systems involving independent atoms exhibit interesting directional properties, there can be similar features created by an open system of atoms in which the atoms are coupled to a common three-dimensional field. It was Dicke~\cite{d54} who pointed out that a collection of a large number of atoms coupled to a common EM field can radiate collectively such that the spontaneous radiation can be enhanced in certain directions. Since then there have been many studies of the collective radiative properties of multi-atom system demonstrating the dependence of the emitted radiation on the number of atoms and the geometry of the emitting system~\cite{Lehmberg,r82,bb84,f86,corr1,corr2,corr3,corr4,Carmichael,mf07,Cirac,Wstate,lz14}.

In this paper, we investigate radiative properties of an open system of a line of few atoms and demonstrate that the dipole-dipole interaction between atoms may lead to a collective ordering of the emission along some specific directions. To determine directions of the emission, we introduce the directivity function of the emitted radiation field and study the dependence of the function on the distance between the atoms. We analyze the directional properties of the radiation field for two different configurations of the atomic chains, one mimicking an atom in front of a mirror and the other an atom inside a cavity. In the first, we choose the left-hand-side atom of the chain as a probe atom and examine conditions under which the system may radiate only to those modes whose propagation vectors lie within a small solid angle about the inter atomic axis and oriented in one of the two directions of the inter atomic axis, either right (forward) or left (backward) direction. Such a system can be regarded as an atom in front of perfectly reflecting or perfectly transmitting atomic mirror. In the second arrangement, we choose the middle atom of the chain as a probe atom and examine conditions under which the system may radiate only to those modes whose propagation vectors lie within a small solid angle about the inter atomic axis. Such a scheme can be regarded as an atom inside a single-mode cavity.

We find that there is a threshold for the inter atomic distances above which a highly directional emission could be achieved. Below the threshold the emission is spherically symmetric. However, we find a population trapping in one of the mirror atoms. For the cavity-like arrangement, the directivity function depends strongly on the number of atoms contained in the chain and the distance between them. For a $3$-atom chain and atomic distances above the threshold, two radiative modes of different spatial directions are available for the emission, one in the direction parallel and the other in the direction normal to the inter-atomic axis. Below the threshold, the system can radiate only to the mode normal to the inter atomic axis. For a $5$-atom chain and distances above the threshold, only one mode is available for emission, either normal or parallel to the atomic line. Thus, there exist ranges of the inter atomic distances under which the atomic chain exhibits features characteristic of a single-mode cavity.

The paper is organized as follows. In Sec.~\ref{sec2}, we describe the master equation of the density operator of the system and the mathematical approach used in the evaluation of the density matrix elements. We introduce the definitions of the directivity function, reflection and transmission coefficients of the radiation field emitted by a chain of atoms. In Sec.~\ref{sec3}, we examine the conditions for the mirror-type behavior of short chains composed of $3$ and $5$ atoms. We observe the transient transfer of the population between the atoms and the transient directivity function for an initial condition in which the probe atom is prepared in its excited state. Then, we examine the directivity function of the stationary field when the probe atom is driven by a continuous wave (cw) laser field. Section~\ref{sec4} is devoted to the problem of a cavity formation with atomic mirrors. We are particularly interested in the possibility of the system to concentrate the radiation along the inter atomic axis and thus to behave as a single-mode cavity. Polar diagrams are given to illustrate the mirror and cavity-like features of atomic chains and to show how the features are sensitive to distances between the atoms.  The results are summarized in Sec.~\ref{sec5}. The paper concludes with an Appendix6A in which we give details of the derivation of the atomic correlation functions in terms of the populations of the collective states of a three-atom system and the coherences between them and Appendix6B where the calculation of equations of motion has been presented.

\section{Radiative properties of a chain of atoms}\label{sec2}

We consider a system composed of $N$ identical two-level atoms located at fixed positions $\vec{r}_{i}$ and coupled to the three-dimensional electromagnetic field whose modes are initially in a vacuum state $\ket{\{0\}}$. Each atom has an excited state $\ket{e_{i}}$ and a ground state $\ket{g_{i}}$ separated by energy~$\hbar\omega_{0}$ and connected by a transition dipole moment~$\vec{\mu}$. 
\begin{figure}[t]
\centering
\includegraphics[width=0.8\columnwidth]{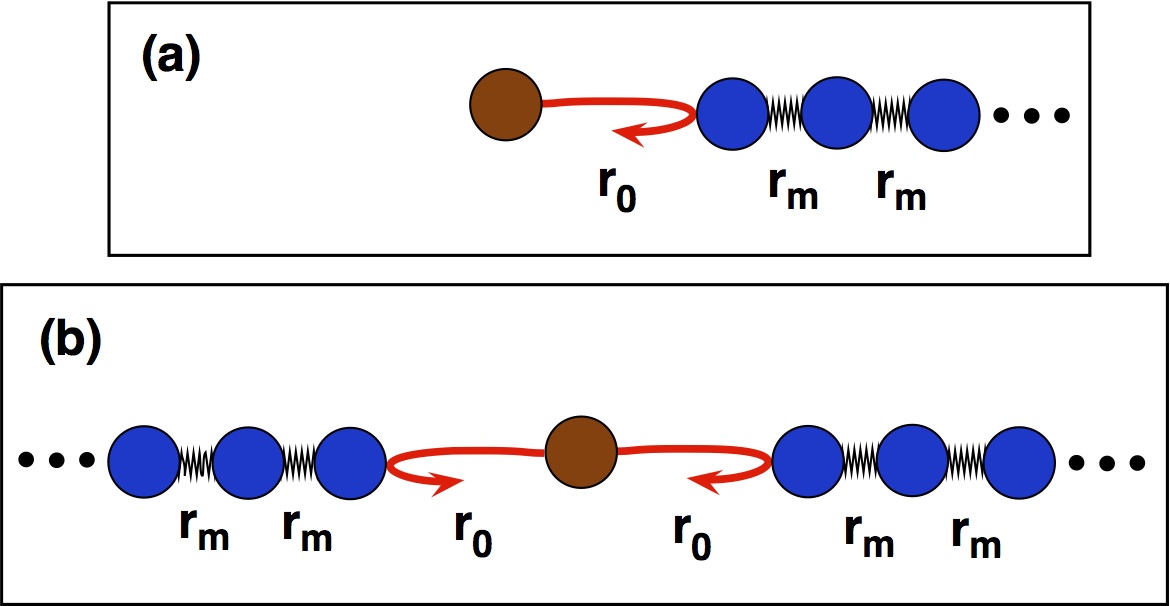}
\caption{(Color online) Two different arrangements of atoms in a line to demonstrate that a chain of closely located and interacting atoms can act as an atomic mirror or cavity. (a) The left-side atom of the chain, specially chosen as a "probe" atom, is located at distance $r_{0}$ from its nearest neighbor with the remaining atoms equally separated from each other by a distance $r_{m}< r_{0}$. (b) The middle atom of the chain,  chosen as a probe atom, is separated from its next-neighbors by $r_{0}$, while the remaining atoms are equally separated by a distance~$r_{m}< r_{0}$. }
\label{fig1}
\end{figure}
The atoms are arranged in a line, and we consider two cases shown in~Fig.~\ref{fig1}. In the first case, illustrated in Fig.~\ref{fig1}(a), we assume that the left-side atom, chosen as a ``probe" atom is separated from its next-neighbor by a distance $r_{0}$ which is larger than the separation $r_{m}$ between the remaining $N-1$ atoms, $r_{0}> r_{m}$. If the probe atom is excited into its upper level, it will spontaneously decay into the ground state emitting the radiation field that can be absorbed by the chain of closely located atoms and then re-emitted by the atoms towards the probe atom. Thus, the chain of closely located atoms could act as a mirror, directing the emitted radiation into a cone about the interatomic axis and turned towards the probe atom. In the second case, illustrated in Fig.~\ref{fig1}(b), we assume that the middle atom of the chain is separated from its adjacent neighbors by a distance $r_{0}$, which is much larger that the separation $r_{m}$ between the remaining atoms. This arrangement may model a situation of an atom located inside a cavity whose mirrors are formed by two chains of equally distant atoms. 

In practice this scheme could be realized by extending the recently demonstrated scheme involving two superconducting qubits coupled to a 1D field~\cite{mlynek, nissen,loo} to the case of three or five qubits coupled to a 2D field. In the experiment of Ref.~\cite{loo}, effective separations of $\lambda$ and $3\lambda/4$ were achieved between the fixed qubits by changing the qubit transition frequencies. 

\subsection{Master equation}

When the system is coupled to a reservoir the state of the total system, the chain of atoms plus the reservoir field, is described by the density operator $\rho_{T}$. The reduced density operator describing the properties of only the chain of atoms is obtained by tracing the total density operator $\rho_{T}$ over the states of the reservoir, $\rho ={\rm Tr}_{R}\rho_{T}$. The master equation describing the time evolution of the reduced density operator has the form~\cite{agarwal,fs,ficek-book}
\begin{align}
\frac{\partial \rho}{\partial t} &= -\frac{i}{\hbar}\left[H_{0}+H_{L}+H_{dd},\rho\right] \nonumber\\
&-\frac{1}{2}\sum_{i=1}^N\gamma \left( \left[S_{i}^{+},S_{i}^{-}\rho\right] + \textrm{H.c.}\right)\nonumber\\
&-\frac{1}{2}\sum_{i\neq j =1}^N\gamma_{ij} \left( \left[S_{i}^{+},S_{j}^{-}\rho\right] + \textrm{H.c.}\right) ,\label{q1}
\end{align}
where $\gamma$ is the spontaneous emission damping rate of the individual  atoms, equal to the Einstein $A$ coefficient, $S_{i}^{+}=\ket{e_{i}}\bra{g_{i}}$ and $S_{i}^{-}=\ket{g_{i}}\bra{e_{i}}$ are the dipole raising and lowering operators of atom $i$, and $\gamma_{ij}$ is the collective damping rate 
\begin{align}
\gamma_{ij} &= \frac{3}{2}\gamma\left\{\left[1 -\left(\hat{\mu}\cdot\hat{r}_{ij}\right)^{2}\right]\frac{\sin\eta_{ij}}{\eta_{ij}}\right. \nonumber\\
&\left. + \left[1-3\left(\hat{\mu}\cdot\hat{r}_{ij}\right)^{2}\right]\left(\frac{\cos\eta_{ij}}{\eta_{ij}^2}-\frac{\sin\eta_{ij}}{\eta_{ij}^3}\right)\right\} , \label{q2}
\end{align}
with
\begin{align}
\eta_{ij} = k\,r_{ij}=2\pi r_{ij}/\lambda ,\quad \vec{r}_{ij} = r_{ij}\hat{r}_{ij}=\vec{r}_{j} -\vec{r}_{i} ,
\end{align}
in which $r_{ij}$ is the distance between atoms $i$ and $j$, $\hat{r}_{ij}$ is the unit vector in the direction $\vec{r}_{ij}$, and $\lambda$ is the resonant wavelength. 

The master equation (\ref{q1}) describes the atomic dynamics under the Born-Markov and rotating-wave approximations~\cite{scullybook}, $H_{0}$ is the Hamiltonian describing the free energy of the atoms 
\begin{align}
H_{0} = \hbar\sum_{i=1}^{N}\omega_{0}S^{+}_{i}S_{i}^{-} ,\label{q3}
\end{align}
$H_{L}$ is the Hamiltonian describing the interaction of the probe atom with an external driving field of frequency $\omega_{L}$:
\begin{align}
H_{L} = \frac{1}{2}\hbar\Omega_{0}\left(S_{1}^{+}e^{-i\omega_{L}t} + S_{1}^{-}e^{i\omega_{L}t}\right) ,\label{q3a}
\end{align}
where $\Omega_{0}$ is the Rabi frequency of the driving field, and $H_{dd}$ is the Hamiltonian describing the dipole-dipole interaction between the atoms
\begin{align}
H_{dd} =  \hbar\sum_{i\neq j=1}^{N}\Omega_{ij}\left(S_{i}^{+}S_{j}^{-} + S_{j}^{+}S_{i}^{-}\right) ,\label{q4}
\end{align} 
where $\Omega_{ij}$ is the dipole-dipole interaction strength between atoms $i$ and $j$, defined by 
\begin{align}
\Omega_{ij} &= \frac{3}{4}\gamma\left\{\left[1-3\left(\hat{\mu}\cdot\hat{r}_{ij}\right)^{2}\right]\left(\frac{\sin\eta_{ij}}{\eta_{ij}^2}+\frac{\cos\eta_{ij}}{\eta_{ij}^3}\right)\right. \nonumber\\
 &\left. -\left[1-\left(\hat{\mu}\cdot\hat{r}_{ij}\right)^{2}\right]\frac{\cos\eta_{ij}}{\eta_{ij}}\right\} .\label{q5}
\end{align}

The parameters $\gamma_{ij}$ and $\Omega_{ij}$ depend on the separation between the atoms. For large separations, $\eta_{ij}\gg 1$, and then both coupling parameters approach zero.  For $\eta_{ij}\ll 1$ the parameter $\gamma_{ij}$ reduces to $\gamma$ while $\Omega_{ij}$ becomes large and strongly dependent on $r_{ij}$. It is well known that $\Omega_{ij}$ plays the important role in the collective behavior of multi-atom systems and we shall see that it has important effect on the distribution of the radiation field emitted by a chain of atoms. The calculation of the equations of motion for atomic populations and coherences for a time-dependent state vector has been outlined briefly in~Appendix6B.

\subsection{Directivity function, reflection and transmission coefficients}\label{sec2a}

The intensity of the radiation field emitted at time~$t$ in the direction specified by the polar angle $\theta$ between the direction of observation $\vec{R}$ and the direction of the atomic axis $\vec{r}_{ij}$ can be expressed in terms of the correlation functions of the atomic dipole operators as
\begin{align}
I(\theta,t) = u(\phi)\sum_{i,j=1}^{N}\gamma \langle S_{i}^{+}(t)S_{j}^{-}(t)\rangle e^{ikr_{ij}\cos\theta} ,\label{q6}
\end{align}
where $u(\phi)=(3/8\pi)\sin^{2}\phi$ is the radiation pattern of a single atomic dipole, with $\phi$ the angle between the observation direction $\vec{R}$ and the direction of the atomic transition dipole moment $\vec{\mu}_{i}$. Since the radiation intensity $I(\theta,t)$ is symmetric about the interatomic axis, it defines a two-dimensional surface called the polar radiation pattern of the emitting system. 

We may introduce the {\it directivity} function determining of how effective the atoms are in converging the emitted radiation into a small solid angle centered about the direction $\theta$. The directivity function $D(\theta,t)$ in the direction $\theta$ at time $t$ is defined as the ratio of the radiation intensity $I(\theta,t)$ emitted in the direction $\theta$ divided by the total radiation intensity $I(t)$:
\begin{align}
D(\theta, t) = \frac{u(\phi)}{I(t)}\sum_{i,j=1}^{N}\gamma \langle S_{i}^{+}(t)S_{j}^{-}(t)\rangle e^{ikr_{ij}\cos\theta}  ,\label{q9}
\end{align}
where the total radiation intensity $I(t)$ at time $t$ is obtained by integrating $I(\theta,t)$ over $\theta$:
\begin{align}
I(t) = \int I(\theta,t)d\theta = \sum_{i,j=1}^{N}\gamma_{ij}\langle S_{i}^{+}(t)S_{j}^{-}(t)\rangle ,\label{q10}
\end{align}
in which $\gamma_{ii}=\gamma_{jj}=\gamma$ and $\gamma_{ij}\, (i\neq j)$ is given in Eq.~(\ref{q2}). The directivity function is a measure of how effective the system is in concentrating the radiation in a given direction. It is equivalent to the probability density of detecting a fluorescence photon traveling in the direction $\theta$.

Since our interest is in situations where a chain of atoms works as an atomic mirror, an important factor is the ability of producing highly directional patterns of the radiation concentrated in one of the two directions along the interatomic axis, either $\theta=0$ or $\theta=\pi$. Following the arrangement illustrated in Fig.~\ref{fig1}(a), $D(\theta=\pi,t)$ describes the radiation field emitted along the atomic axis in the direction towards the probe atom, the "backward" direction. Thus, it would correspond to the reflection coefficient. On the other hand, the directivity $D(\theta=0,t)$ describes  the radiation field emitted along the atomic axis in the direction away from the probe atom, the ``forward" direction. Therefore, it would correspond to the transmission coefficient of the atomic mirror. 

Thus, we may define the {\it reflection} coefficient of the chain of atoms as the ratio of the radiation 
intensity emitted in the direction $\theta= \pi$ to the total radiation intensity 
\begin{align}
R(t) \equiv D(\theta\!=\!\pi, t) = \frac{u(\phi)}{I(t)}\!\sum_{i,j=1}^{N}\!\gamma \langle S_{i}^{+}(t)S_{j}^{-}(t)\rangle e^{ikr_{ij}} .\label{q12}
\end{align}
The reflection coefficient is a measure of how effective the atoms are in concentrating the radiation about one side of the interatomic axis, i.e. about the direction $\theta =\pi$. The coefficient $R(t)$ is equivalent to the probability density of detecting a fluorescence photon traveling in the direction~$\theta =\pi$. 

Similarly, we can define the {\it transmission} coefficient of the chain as the ratio of the radiation intensity emitted in the direction $\theta=0$ to the total intensity of the field emitted 
\begin{align}
T(t) \equiv D(\theta\!=\!0,t) = \frac{u(\phi)}{I(t)}\!\sum_{i,j=1}^{N}\!\gamma \langle S_{i}^{+}(t)S_{j}^{-}(t)\rangle e^{-ikr_{ij}} .\label{q11}
\end{align}
Obviously, $T(t)=1$ would correspond to complete transmission whereas $R(t)=1$ would correspond to complete reflection of the radiation field emitted along the atomic axis.

\subsection{Directional properties of the radiation field}\label{sec2b}

The quantity of central interest is the directivity function which can be determined from analyzing the polar radiation pattern of a chain of atoms. We may determine general conditions under which 
the radiation pattern of $N$ atoms would be highly non spherical and its maximum is concentrated along the inter atomic axis. The conclusions will serve 
as reference for choosing distances between the atoms and for calculations of the atomic populations and correlations.  

The expression (\ref{q6}) can be written as a sum of $N$ terms
\begin{align}
I(\theta,t) = \sum_{i<j=1}^{N}I_{ij}(\theta,t) ,\label{q7}
\end{align}
where
\begin{align}
I_{ij}(\theta,t) &= u(\phi)\gamma \left\{\frac{1}{N\!-\!1}\!\left(\langle S_{i}^{+}(t)S_{i}^{-}(t)\rangle\!+\!\langle S_{j}^{+}(t)S_{j}^{-}(t)\rangle\right)\right. \nonumber\\
&\left. +\, 2{\rm Re}\{\langle S_{i}^{+}(t)S_{j}^{-}(t)\rangle\}\cos\left(kr_{ij}\cos\theta\right)\right. \nonumber\\
&\left. -\, 2{\rm Im}\{\langle S_{i}^{+}(t)S_{j}^{-}(t)\rangle\}\sin\left(kr_{ij}\cos\theta\right)\right\} .\label{q8}
\end{align}
We see that the contribution of the atoms to the intensity occurs in pairs of different combinations of the atoms. Therefore, the system of radiating atoms can be considered as made up of a number of short two-atom elements and the total intensity is obtained by summing up the intensities of the fields produced by all the elements.  

If we wish a short chain of atoms to work like a mirror with a large convergence and reflectivity of the radiation emitted by a probe atom located at either end of the chain, we should arrange the atoms such that the total field emitted (scattered) could be highly focused along the interatomic axis with a pronounced maximum in the backward direction $\theta = \pi$ and a minimum, preferably zero emission in the forward direction $\theta = 0$ with respect to the line center. In order to find the conditions for concentrating the radiation in the direction $\theta= \pi$, let us examine the intensity (\ref{q8}) in more details. 

From Eq.~(\ref{q8}) it is seen that there are three terms determining the radiation pattern.
The first term in Eq.~(\ref{q8}) is just the sum of the populations of the two atoms involved, the probabilities that the atoms are in their excited states. This term is independent of $\theta$ and therefore contributes uniformly in all directions. The second term depends on $\theta$ and varies as $\cos\left(kr_{ij}\cos\theta\right)$ with an amplitude equal to the real part of atomic correlations. Consequently, this term could contribute to the radiation pattern only if the correlations between the atoms would have nonzero real part, ${\rm Re}\{\langle S_{i}^{+}S_{j}^{-}\rangle\}\neq 0$. However, the cosine term, although dependent on $\theta$,  would produce intensity maxima in the $\theta =0, \pi/2, 3\pi/2$ and $\pi$ directions. As such this term is not effective in concentrating the radiation along the interatomic axis. This simple argument suggests that the atoms should be arranged such that the cosine term vanishes. This could be achieved when the correlations between the atoms have zero real part. 

The third term contributing to the radiation intensity~(\ref{q8}) varies as $\sin\left(kr_{ij}\cos\theta\right)$ and hence can affect the radiation intensity in a decidedly different way than the cosine term.  
An important difference is that $\sin\left(kr_{ij}\cos\theta\right)$ vanishes for $\theta =\pi/2$ and $3\pi/2$. This means that the sine term does not contribute to the radiation emitted in the direction perpendicular to the interatomic axis. Consequently, an appreciable concentration of the radiation could be achieved along the interatomic axis by choosing proper distances $r_{ij}$ between the atoms at which $\sin\left(kr_{ij}\right)=\pm 1$. Furthermore, since the sine is an odd function, it follows that $\sin\left(kr_{ij}\cos 0^{\circ}\right)= -\sin\left(kr_{ij}\cos 180^{\circ}\right)$, which means that independent of the separation between the atoms a maximum in the backward direction is always accompanied by a minimum in the forward direction.  

Obviously, the sine term could influence the angular distribution of the radiation intensity only if the atomic correlations would have nonzero imaginary parts, ${\rm Im}\{\langle S_{i}^{+}S_{j}^{-}\rangle\}\neq 0$. In addition, the sign of the imaginary part of the atomic correlations dictates the choice of distances between the atoms, at which the emission would be enhanced in the backward direction (high reflection) and reduced in the forward direction (low transmission). Thus, if ${\rm Im}\{\langle S_{i}^{+}S_{j}^{-}\rangle\} > 0$, the sine term will display a maximum in the backward direction for atomic distances at which $\sin(kr_{ij})=1$. This condition is satisfied when the atomic separations are
\begin{align}
r_{ij} = \frac{1}{4}(2n+1)\lambda ,\quad n\in\{0,2,4,\ldots \}
\end{align}
However, if the coefficient of the sine term is negative, ${\rm Im}\{\langle S_{i}^{+}S_{j}^{-}\rangle\} < 0$, a different choice of the distances is required at which $\sin(kr_{ij})=-1$. This condition is satisfied when
\begin{align}
r_{ij} = \frac{3}{4}(2n+1)\lambda ,\quad n\in\{0,2,4,\ldots \}
\end{align}
We see that there is a lower bound imposed on the distances between the atoms, either $r_{ij}=\lambda/4$ or $r_{ij}=3\lambda/4$, above which a one-sided emission along the interatomic axis can be achieved,  i.e., a maximum of radiation in the direction $\theta = 0$. For distances smaller that the lower bound, the one-sided emission is expected to be significantly reduced.

In the following, we limit ourselves to short chains containing only $N=3$ and $N=5$ atoms. Also in the absence of the driving field, $\Omega_{0}=0$, the $N=3$ case can be solved in closed form yielding simple expressions. In the~Appendix6A, we outline the calculation of atomic correlation functions in the collective state basis.

\subsection{Mathematical approach}

To study the radiative behavior of a chain of interacting atoms, we require the time evolution of the populations of the atoms and the coherence between them. These are given by the diagonal and off-diagonal density matrix elements, respectively. If the space of the atomic system is spanned in the basis of the eigenstates of the free Hamiltonian $H_{0}$, we readily find that the basis is composed of $2^{N}$ state vectors, i.e. for a chain composed of $N=3$ atoms, the basis is composed of $8$ vectors $\ket{i_{1}j_{2}k_{3}}$, whereas for $N=5$ atoms it is composed of $64$ vectors $\ket{i_{1}j_{2}k_{3}l_{4}m_{5}},\, \{i,j,k,l,m\} \in\{ g,e\}$.

Hence, in the simplest case of $N=3$ atoms, the master equation (\ref{q1}) provides us with a system of $64$ coupled linear equations to be solved, in principle, a $63\times 63$ matrix to be diagonalized. For $N=5$ we get a system of $4096$ coupled linear equations whose solution requires a $4095\times 4095$ matrix to be diagonalized.
Needless to say, it is not easily accomplished. Therefore, we shall use numerical methods. To get solutions for the density matrix elements, the following approach will be taken.

From the master equation (\ref{q1}), we find equations of motion for the density matrix elements, which can be written in a matrix form as
\begin{equation}
\dot{\vec{X}}\left( t\right) = {\cal A} \vec{X}\left( t\right) +\vec{R},  \label{q17}
\end{equation}
where $\vec{X}\left( t\right) $ is a column vector composed of the density matrix elements, $\vec{R}$ is a column vector composed of the inhomogenous terms, and ${\cal A}$ 
is a matrix of the coefficients appearing in the equations of motion of the density matrix elements. 

A direct integration of Eq.~(\ref{q17}) leads to the following formal solution for $\vec{X}\left( t\right)$
\begin{align}
\vec{X}\left( t\right) =\vec{X}\left( t_{0}\right) e^{{\cal A}t}-\left(1-e^{{\cal A}t}\right) {\cal A}^{-1}\vec{R},  \label{q18}
\end{align}
where $t_{0}$ is the initial time.

Because the determinant of the matrix ${\cal A}$ is different from zero, there exists a complex invertible matrix ${\cal U}$ which diagonalizes ${\cal A}$, and $w = {\cal U}^{-1}{\cal A}{\cal U}$ is the diagonal matrix of complex eigenvalues. By introducing new vectors $\vec{Y}={\cal U}^{-1}\vec{X}$ and $\vec{T}={\cal U}^{-1}\vec{R}$, we can rewrite (\ref{q18}) as
\begin{equation}
\vec{Y}\left( t\right) =\vec{Y}\left( t_{0}\right) e^{w t}-\left(1-e^{w t}\right) w^{-1}\vec{T} ,  \label{q19}
\end{equation}
or in component form
\begin{equation}
Y_{n}\left( t\right) =Y_{n}\left( t_{0}\right) e^{w_{n}t}-\sum_{m=1}^{s}\!\left( w^{-1}\right)_{nm}\!\left( 1-e^{w_{m}t}\right) T_{m} ,  \label{q20}
\end{equation}
in which $s=63$ for the case of $N=3$ atoms and $s=4095$ for $N=5$ atoms. To obtain solutions for $X_{n}\left( t\right) $ we must determine the eigenvalues $w_{n}$ and eigenvectors $Y_{n}\left( t\right) ,$ which are readily evaluated by a numerical diagonalization of the matrix ${\cal A}$.

The steady-state values of the density matrix elements can be found from Eq.~(\ref{q20}) by taking $t\rightarrow \infty $, or more directly by setting the
left-hand side of Eq.~(\ref{q17}) equal to zero, and then
\begin{equation}
\vec{X}\left( \infty \right) =-{\cal A}^{-1}\vec{R},  \label{q21}
\end{equation}
or in component form
\begin{equation}
X_{n}\left( \infty \right) =-\sum_{m=1}^{s}\left( {\cal A}^{-1}\right)_{nm}R_{m} .  \label{q22}
\end{equation}

In what follows, we shall use the solutions (\ref{q20}) and (\ref{q22}) to illustrate the radiative properties of the atomic chains. In particular, we calculate the transient populations of the atoms and the steady-state directivity function.

\section{Radiating atom at front of an atomic mirror}\label{sec3}

Consider first the configuration illustrated in Fig.~\ref{fig1}(a), a single atom distance $r_{12}$ from a finite-size chain of closely located atoms. This arrangement could constitute a radiating atom at front of an atomic mirror. In order to find the mirror-like characteristics of the chain that could be inferred from the radiative properties of the systems, we examine the directivity function of the radiation field emitted by the system. The directivity function, which is determined by the angular distribution of the radiation intensity, Eq.~(\ref{q7}), depends on temporal and spatial factors, the time of the evolution of the system and the separation between the atoms. It is clear from Eq.~(\ref{q7}) that in general the temporal and spatial factors cannot be separated. The angular distribution of the radiation field at a given time can be different for different separations between the atoms. Moreover, the transient radiation intensity is a sensitive function of the initial atomic conditions. Even at the initial time $t=0$ the angular distribution of the radiation intensity can depend on the separation $r_{ij}$ if the system was prepared in a state with nonzero interatomic correlations, either ${\rm Re}\{\langle S_{i}^{+}S_{j}^{-}\rangle\}$ or ${\rm Im}\{\langle S_{i}^{+}S_{j}^{-}\rangle\}$ different from zero. In order to study the angular distribution of the radiation field,  it is important to understand the radiative behavior of individual atoms in the chain. Therefore, we first consider the evolution of the populations of the atoms. Following this discussion, we display the variation of the directivity function with time for different distances between the atoms. We assume that the probe atom is initially excited and separated from the front atom of the ``mirror", its nearest neighbor, at a distance much larger than the separation between the ``mirror" atoms. The method of preparing the probe atom in the excited state will not concern us but it might be done with a short laser pulse, for example.

\subsection{Transient transfer of the population}\label{sec3a}

Our interest is in the time evolution of the directivity function of the emitted field for an initially excited probe atom, the case illustrated in Fig.~\ref{fig1}(a). In order to study this evolution, we first look at the time evolution of the populations of the atoms for chains composed of $N=3$ and $N=5$ atoms. The aim is to first determine and then optimize the conditions which not only allow the radiation emitted by the probe atom to be emitted back (reflected) to it but also to achieve a strong convergence of the radiated field along the interatomic axis. 

\begin{figure}[h]
\includegraphics[width=4.55cm]{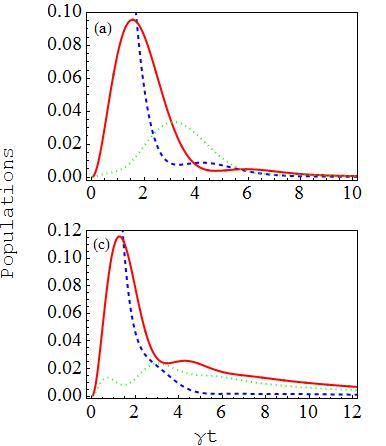}
\hspace{-0.4cm}
\includegraphics[width=4.1cm]{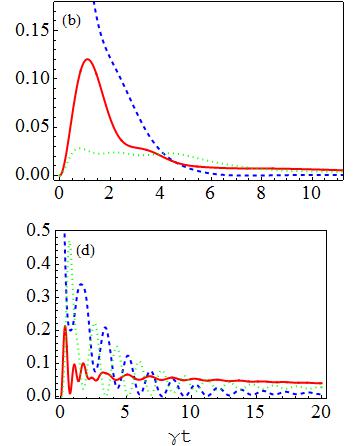}
\caption{(Color online) Time evolution of the atomic populations $\rho_{11}(t)$ (dashed blue line), $\rho_{22}(t)$ (solid red line), $\rho_{33}(t)$ (dotted green line) plotted for the case of an initially excited probe atom $1$ located in front of a line of two atoms, $2$ and $3$, and several different sets of distances between the atoms: (a) $(r_{12}, r_{23}) =(\lambda/2,\lambda/4)$, (b) $(r_{12}, r_{23}) =(\lambda/4,\lambda/6)$, (c) $(r_{12}, r_{23}) =(\lambda/3,\lambda/5)$, and (d) $(r_{12}, r_{23}) =(\lambda/8,\lambda/9)$.}
\label{fig2}
\end{figure}

\begin{figure}[h]
\includegraphics[width=4.38cm]{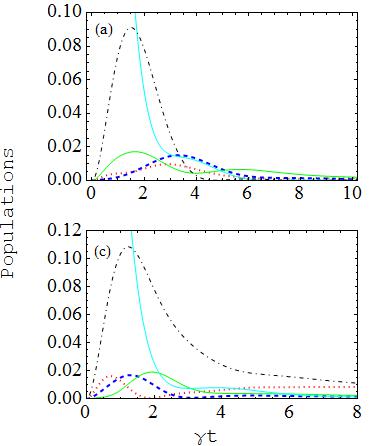}
\includegraphics[width=4.1cm]{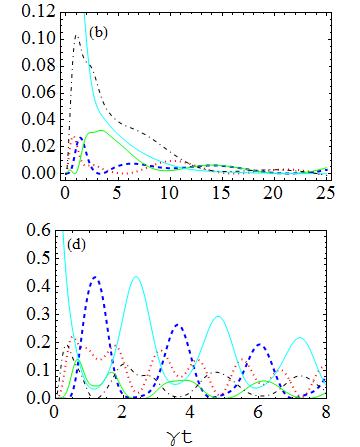}
\caption{(Color online) Time evolution of the atomic populations, $\rho_{11}$ (cyan solid line), $\rho_{22}$ (black dashed-dotted line), $\rho_{33}$ (red dotted line), $\rho_{44}$ (green solid line), and $\rho_{55}$ (blue dashed line)  plotted for the case of an initially excited probe atom $1$ located in front of a line of four atoms and several different sets of distances between the atoms: (a) $(r_{12}, r_{i,i+1}) = (\lambda/2,\lambda/4)$, (b) $(r_{12}, r_{i,i+1}) =(\lambda/4,\lambda/6)$, (c) $(r_{12}, r_{i,i+1}) =(\lambda/3,\lambda/5)$, and (d) $(r_{12}, r_{i,i+1}) = (\lambda/8,\lambda/9)$.}
\label{fig3}
\end{figure}

We evaluate the transient populations of the atoms assuming there is no driving field $(\Omega_{0}=0)$ and that initially the probe atom was in its excited state.
The corresponding results for the time evolution of the atomic populations for different distances between the atoms are presented in~Figs.~\ref{fig2} and~\ref{fig3}. 

Figure~\ref{fig2} shows the transient transfer of the population between $N=3$ atoms; the initially excited probe atom and a group of two atoms forming a mirror. The populations are computed for several different sets of distances $(r_{12}, r_{23})$ between the atoms. For distances $r_{ij}>\lambda/5$ the initial population of the probe atom decays exponentially in time whereas the populations of the mirror atoms $2$ and $3$ build up with small oscillations. However, there is no tendency of the mirror atoms to transfer their populations back to the probe atom. A small oscillatory behavior of the population $\rho_{11}(t)$ can be seen for short times, $t<5/\gamma$, but it is not periodic. A periodic oscillatory behavior is observed for the populations of atoms $2$ and $3$ that the mirror atoms periodically exchange their populations without transferring it to the probe atom. The reason is that the distance $r_{23}$ is much smaller than $r_{12}$, resulting in a stronger dipole-dipole interaction between the mirror atoms, $\Omega_{23}\gg \Omega_{12}$.
For smaller distances, the population of the probe atom begins to show periodic oscillations, with the population actually oscillating between the probe atom and only the rear atom of the mirror. Except for very short times, the population of the middle atom $2$ remains almost constant during the evolution of the system.
Note that the transfer of the population between the atoms is not complete since the populations remain nonzero, $\rho_{ii}(t)\neq 0$ for all $t > 0$. 
It is interesting that atom $2$, the front atom of the mirror, appears as an mediator in the population exchange between the probe atom and the rear atom of the mirror. In other words, the front atom of the mirror  dictates the direction of the emission. Despite of the possibility to emit in any spatial direction, the probe atom ``prefers" to radiate towards the atomic mirror and vice versa, the mirror atoms prefer to radiate towards the probe atom. 

We now consider a chain containing $N=5$ atoms with the probe atom initially excited and located at a distance $r_{12}$ from the front of a line of four equally separated atoms.
The transient transfer of the population between the atoms for several different sets of distances, $(r_{12}, r_{i,i+1}), \, i \in \{2,3,4 \}$ is shown in Fig.~\ref{fig3}. For large distances the situation is similar to that of three atoms, the population rapidly escapes from the system leaving the atoms unpopulated over a short evolution time. 
For small distances, we observe periodic oscillations of the populations with the periodicity, as before for three atoms, determined by the dipole-dipole interaction strength. However, unlike the three atoms case, the population can be completely transferred from the mirror atoms back to the probe atom such that the mirror atoms are unpopulated at times when the population of the probe atom is maximal. This is a substantial difference compared to the case of $N=3$ atoms where a part of the population was trapped by one of the mirror atoms. Clearly, longer chains are more effective in the complete transfer of an excitation from the probe atom to the mirror atoms and vice versa.

\subsection{Transient directivity function}\label{sec3b}

The periodic exchange of the populations between the atoms, in particular, between the atoms forming an atomic mirror, can lead to the emission of the atoms into certain preferred directions. In order to demonstrate this behavior, we examine the time variation of the directivity function $D(\theta,t)$ of the field radiated by the group of mirror atoms only. Of course it might be argued that it could be hard for experiments to detect the field radiated from a fraction of atoms, but we would like to see under which conditions the field radiated by the group of mirror atoms could be concentrated in the preferred direction along the interatomic axis, either $\theta=0$ or $\theta =\pi$. With a high concentration of the radiated field into a small solid angle centered around $\theta=0$, the atoms could be regarded as a highly reflecting mirror scattering the field back towards the probe atom. 
\begin{figure}[h]
\includegraphics[width=4.27cm]{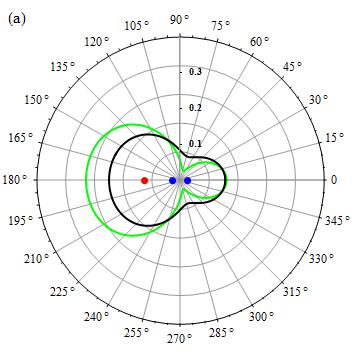}
\includegraphics[width=4.27cm]{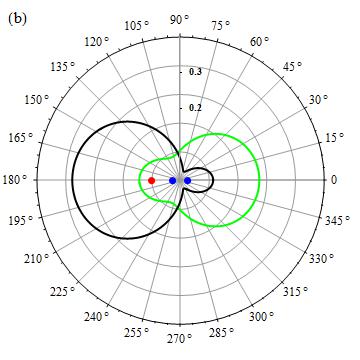}
\includegraphics[width=4.27cm]{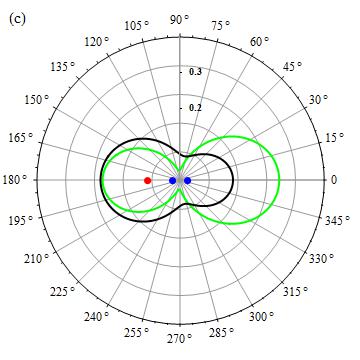}
\includegraphics[width=4.27cm]{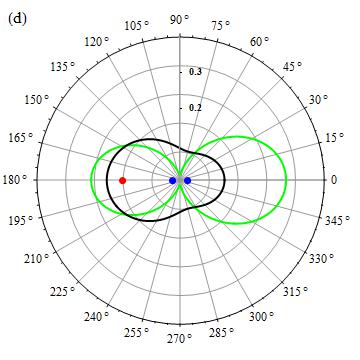}
\caption{(Color online) Polar diagram of the directivity function $D(\theta,t)$ describing the concentration of the field emitted by two atoms forming a mirror when an initially excited probe atom is at distance $r_{12}=\lambda/2$ from the front of the mirror atoms. Distances between the mirror atoms are: in (a) $r_{23}=\lambda/3$, (b) $r_{23}=\lambda/4$, (c) $r_{23}=\lambda/5$, and in (d) $r_{23}=\lambda/8$. Green (black) curve corresponds to a minimum (maximum) in $\rho_{11}(t)$. The time instants for the green and black curves are: (a) $t\in\{3\gamma^{-1}, 4\gamma^{-1} \}$, (b) $t\in\{5\gamma^{-1}, 3.3\gamma^{-1} \}$, (c) $t\in\{9\gamma^{-1}, 3.6\gamma^{-1} \}$,  (d) $t\in\{9.4\gamma^{-1}, 6\gamma^{-1} \}$, respectively.  }
\label{fig4}
\end{figure}

The directionality function of the field emitted by two (mirror) atoms and detected on a sphere around the line of three atoms is shown in Fig.~\ref{fig4}. We keep the distance of the probe atom from the front of the mirror atoms at $r_{12}=\lambda/2$ and consider the variation of $D(\theta,t)$ with the distance between the mirror atoms at two different times. It is seen that the emission is concentrated along the interatomic axis. For a distance between the mirror atoms $r_{23}=\lambda/4$, the directivity function is significantly enhanced in one direction, that the emission is almost one-sided either in $\theta=0$ or $\theta=\pi$. The directivity function oscillates in time in such a manner that it develops a maximum in the direction $\theta =0$ at later times in which it was a minimum at earlier times and vice versa. The oscillation of the directivity results from the oscillation of the population between the mirror atoms, as we have see above in Fig.~\ref{fig2}. The beam width or equivalently the solid angle inside which the emitted field is concentrated remains almost constant.
\begin{figure}[h]
\includegraphics[width=4.28cm]{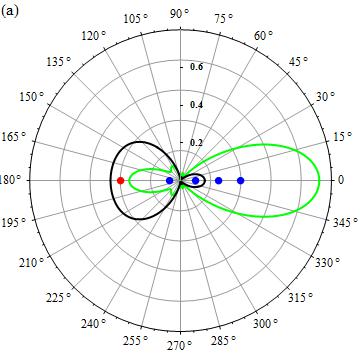}
\includegraphics[width=4.28cm]{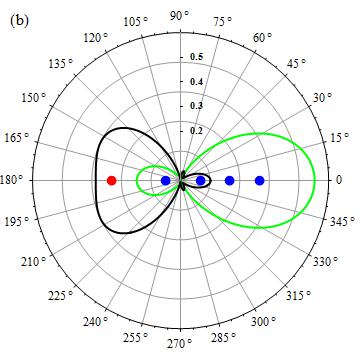}
\includegraphics[width=4.28cm]{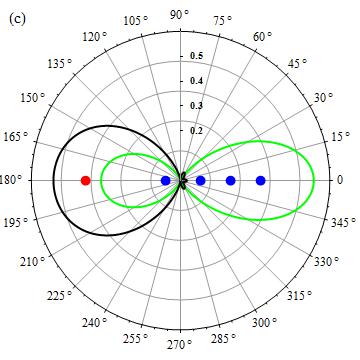}
\includegraphics[width=4.2cm]{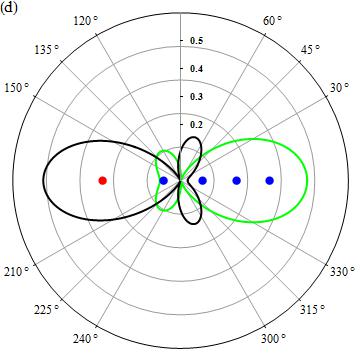}
\caption{(Color online) Polar diagram of the directivity function $D(\theta,t)$ describing the concentration of the field emitted by four atoms forming a mirror when an initially excited probe atom is at distance $r_{12}=\lambda/2$ from the front of the mirror atoms. Distances between the mirror atoms are: in (a) $r_{i,i+1}=\lambda/3$, (b) $r_{i,i+1}=\lambda/4$, (c) $r_{i,i+1}=\lambda/5$, and in (d) $r_{i,i+1}=\lambda/8$.  Green (black) curve corresponds to a minimum (maximum) in $\rho_{11}(t)$. The time instants for the green and black curves are: (a) $t\in\{7.76\gamma^{-1}, 3.7\gamma^{-1} \}$, (b) $t\in\{6.8\gamma^{-1}, 3.4\gamma^{-1} \}$, (c) $t\in\{7\gamma^{-1}, 3.715\gamma^{-1} \}$,  (d) $t\in\{7.6\gamma^{-1}, 9.1\gamma^{-1} \}$, respectively. }
\label{fig5}
\end{figure}

One can also notice from Fig.~\ref{fig4} that the ability of the coupled atoms to concentrate the radiation in one direction decreases with a decreasing distance between the atoms. It is clearly seen that for distances $r_{23}\leq \lambda/8$ the directivity function is almost symmetrical along the interatomic axis. This result is consistent with the general property of the radiation pattern discussed in Sec.~\ref{sec2a} that there is a minimal value of the distance between the atoms $(r_{ij}=\lambda/4)$ above which the sine term can be maximal in the $\theta=0$ direction. For distances smaller than the minimal value the contribution of the sine term is necessarily smaller resulting in the reduction of the ability of the system to concentrate the emission in one direction.

The one-sided emission and then the reflectivity coefficient can be enhanced by increasing the number of atoms forming the atomic mirror. This is demonstrated in Fig.~\ref{fig5} which shows the polar diagram of the directivity function $D(\theta,t)$ for the case of four atoms forming the atomic mirror. As above for three atoms, we keep the probe atom at a fixed distance from the front of the atomic mirror, $r_{12}=\lambda/2$, and consider the variation of $D(\theta,t)$ with the distance between the mirror atoms. It can be seen that similar to the case of two atoms the emission is concentrated mainly along the interatomic axis. The one-sided emission is significantly enhanced and persists even for small distances. Small peaks can be seen at about $2\pi/3$ and $4\pi/3$ degrees. It is easy to verify that the function $\sin\left(kr_{ij}\cos\theta\right)$ when evaluated for $r_{25}=3\lambda/8$, corresponding to the distance between the front and rear atoms of the mirror, attains its maximal value of $\sin\left(kr_{ij}\cos\theta\right)= 1$ for $\cos\theta= 2/3$, which corresponds to directions $\theta \approx 130^{\circ}$.

\subsection{Directivity function of the stationary field}\label{sec3c}

In practice a photo detector located at some position $\vec{R}$ in the far field zone of the radiation field emitted by the atoms would detect the field emitted by the entire set of atoms rather than a fraction of selected atoms only. It is a consequence of the fact that the fields from the probe atom and the mirror atoms are unresolved at the detector. Therefore, we now consider the directivity function of the field radiated by the entire system of atoms including the field radiated from the probe atom. Moreover, we assume that initially all atoms were in their ground states and then the probe atom was exposed to the incident weak laser light. We shall look into the radiation pattern of the emitted field in the steady-state limit, $t\rightarrow \infty$, and compute the directivity function  $D(\theta) \equiv \lim_{t\rightarrow\infty}D(\theta, t)$ for equally distant as well as for non-equally distant $N=3$ and $N=5$ atoms.
Since the directivity function involves the contribution from the populations of the atoms, that a significant directionality can be obtained when the populations are small, we shall assume that the driving field is weak so that  the Rabi frequency $\Omega$ much smaller than $\gamma$.
\begin{center}
\begin{figure}[h]
\includegraphics[width=4.27cm]{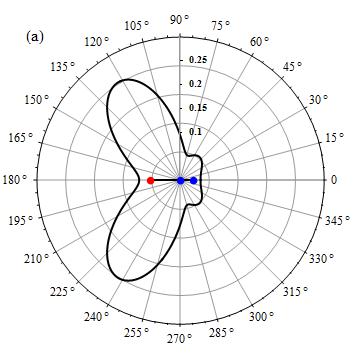}
\includegraphics[width=4.27cm]{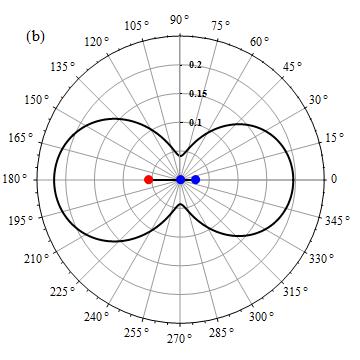}
\includegraphics[width=4.27cm]{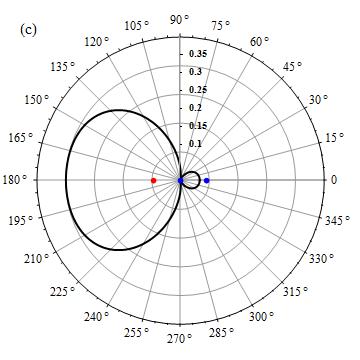}
\includegraphics[width=4.27cm]{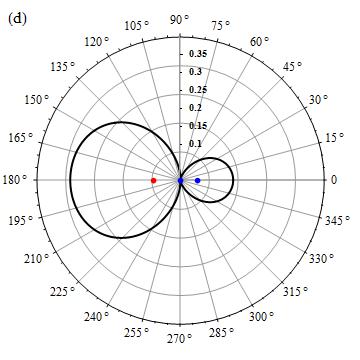}
\caption{(Color online) Polar diagram of the directivity function $D(\theta)$ of the stationary field emitted by three atoms in a line. The left-side atom of the line, which constitutes a probe atom is driven by a cw laser field of the Rabi frequency $\Omega_{0}=0.01\gamma$. In frame (a) $r_{12}=\lambda/2, r_{23}=\lambda/4$.  In frame (b) $r_{12}=\lambda/4, r_{23}=\lambda/8$. In frame (c) $r_{12}= r_{23} = \lambda/4$, and in frame (d) $r_{12}=\lambda/4, r_{23}=\lambda/6$.\\
 }
\label{fig6}
\end{figure}
\end{center}

The directivity function for the radiation field emitted by $N=3$ atoms is shown in Fig.~\ref{fig6}, where frame (a) is for equally distant atoms, while frame (b) is for unequally distant atoms. 
Unfortunately, as one can see from the figure, in both cases that the stationary field radiated by the system is concentrated in directions $\theta=\pi/3$ and $\theta = 5\pi/3$, which significantly departure from the direction of the interatomic axis. 
Therefore, the system of three equally or unequally separated atoms with the continuously driven probe atom cannot be regarded as suitable for the mirror-type behavior of the maximal directivity along the interatomic axis.
\begin{figure}[h]
\includegraphics[width=4.27cm]{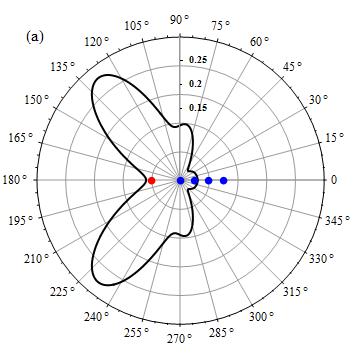}
\includegraphics[width=4.27cm]{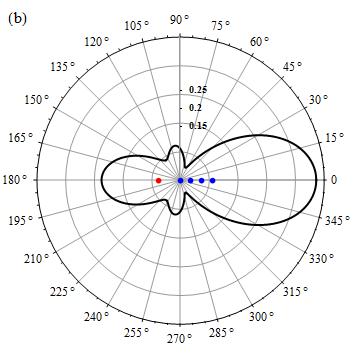}
\includegraphics[width=4.27cm]{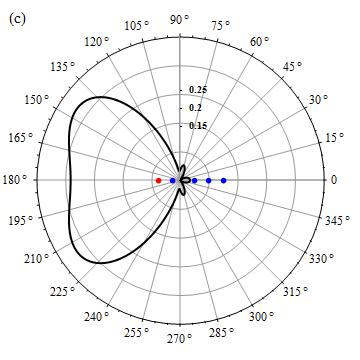}
\includegraphics[width=4.27cm]{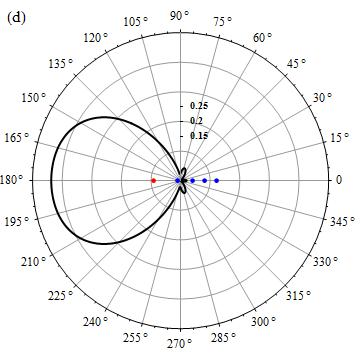}
\caption{(Color online) Polar diagram of the directivity function $D(\theta)$ of the stationary field radiated by a line of $5$ atoms. The left-side atom of the line, which constitutes a probe atom is driven by a cw laser field of the Rabi frequency $\Omega_{0}=0.01\gamma$. In frame (a), the atoms are unequally separated with $r_{12}=\lambda/2$, and $r_{23}=r_{34}=r_{45}=\lambda/4$.  In frame (b), the atoms are unequally separated with $r_{12}=\lambda/4$, and $r_{23}=r_{34}=r_{45}=\lambda/8$. In frame (c), the atoms are equally separated with $r_{12}= r_{23} = r_{34}=r_{45}= \lambda/4$, and in frame (d), the atoms are unequally separated with $r_{12}=\lambda/4$, and $r_{23}=r_{34}=r_{45}=\lambda/6$.}
\label{fig7}
\end{figure}

Figure~\ref{fig7} shows the corresponding situation for the case of $N=5$ atoms. It is seen that in the case of four atoms forming the mirror the directivity function depends crucially on the separation between the atoms. In particular when the distance of the probe atom from the front of the mirror atoms is $r_{12}=\lambda/4$, the radiated field modes available for the emission are only those contained inside a cone centered about the direction $\theta =\pi$. Thus, at that particular distance the chain of four undriven atoms acts as a perfectly reflecting mirror by ``pushing" the radiation in the backward direction.
It can be seen that the optimal conditions for the one-sided emission with a maximum in the direction $\theta =\pi$ are achieved when the probe atom is separated from the front of the mirror atoms by $r_{12}=\lambda/4$ and the mirror atoms are themselves separated by $r_{ij}=\lambda/6$. In this case, the directivity function is nonzero only in directions contained inside a cone limited by $\theta\leq \pm\pi/3$. Thus, the emission is entirely one-sided with no radiation in the $\theta=0$ direction. This means that the transmission coefficient $T=0$, therefore this type of behavior can be regarded as a mirror type with perfect reflectivity. Of course, the reflectivity is accompanied by losses in the sense that the radiation is emitted into a cone of a finite solid angle $2\theta$ determining the beamwidth of the radiated field. The solid angle subtended by the cone seen in Fig.~\ref{fig7}(c) is $2\theta = 120^{\circ}$.

\begin{figure}[h]
\includegraphics[width=4.27cm]{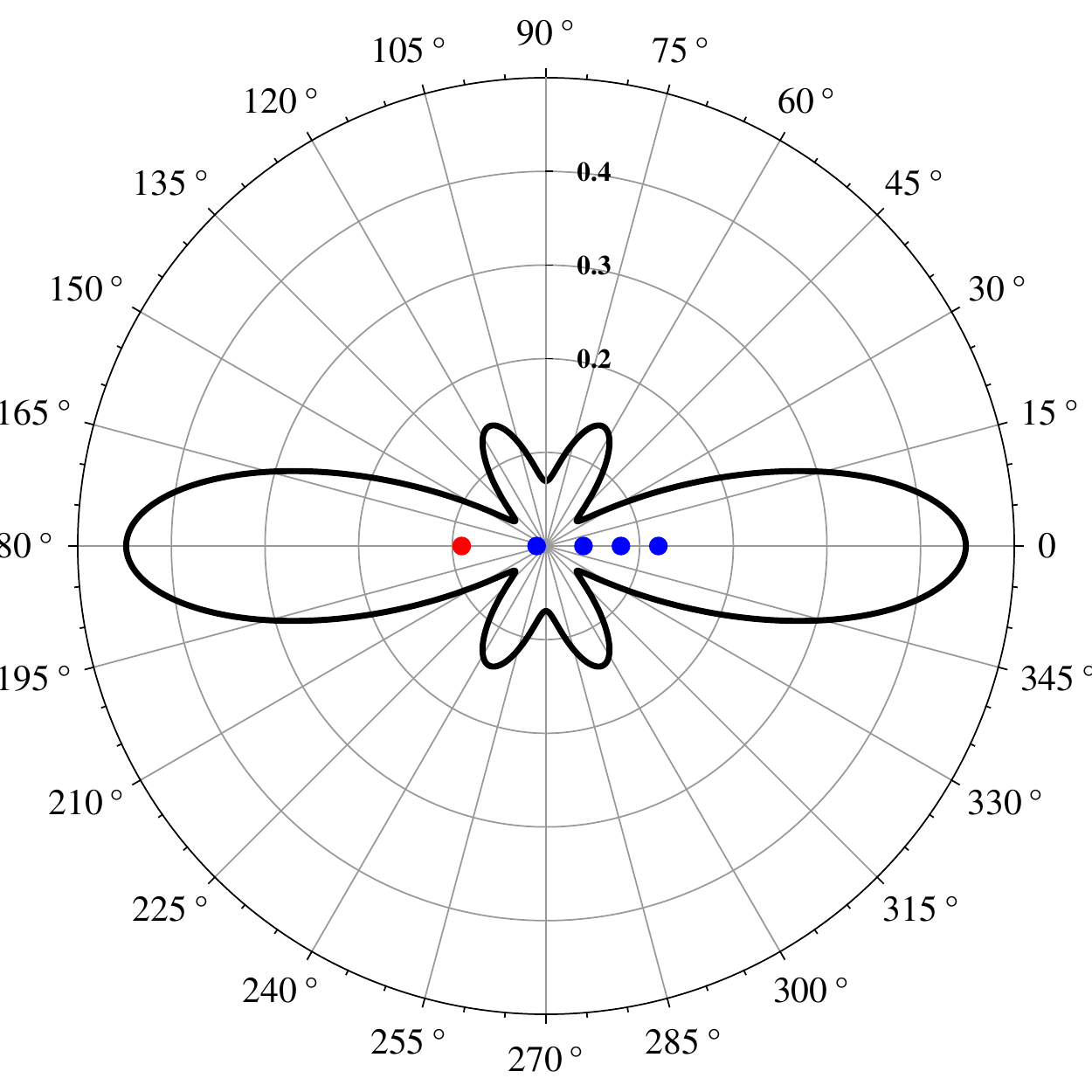}
\caption{(Color online) Polar diagram of the directivity function $D(\theta)$ of the stationary field radiated by a line of $5$ independent atoms $(\Omega_{ij}=0)$. The atoms are unequally separated with $r_{12}=\lambda/4$, and $r_{23}=r_{34}=r_{45}=\lambda/6$.}
\label{fig7a}
\end{figure}

It was pointed out in~\cite{Carmichael} that a strong directivity of the emitted radiation along the interatomic axis can be obtained in a chain of independent atoms, i.e. in the absence of the dipole-dipole interaction. However, the one sided emission seen in frames (c) and (d) of Fig.~\ref{fig7} requires a non-zero dipole-dipole interaction. This is illustrated in Fig.~\ref{fig7a} which shows the directivity function for the same parameters as in Fig.~\ref{fig7}(d) but for independent atoms. Clearly, in the absence of the dipole-dipole interaction between the atoms, the emission is strongly directional, but is symmetrical in the $\theta=0$ and $\pi$ directions. 


In summary of this section, we have seen that the directivity of the emission and the mirror like behavior of a line of atoms depends on the distance between the atoms and the nature of the excitation. At particular distances between the atoms, the system can radiate along the interatomic axis in only one direction $\theta=\pi$, which can be interpreted as a perfect reflection of the radiation towards the probe atom.

\section{Cavity formation with atomic mirrors}\label{sec4}

We now wish to create a cavity with atomic mirrors. For this purpose, we place a probe atom between a pair of finite-size chains of closely located atoms, as illustrated in Fig.~\ref{fig1}(b). These would constitute a radiating atom located inside a one-dimensional cavity. We determine the parameter ranges in which the two chains of $(N-1)/2$ atoms could act as mirrors to the field emitted by the probe atom. We examine two systems containing different numbers of atoms. First, we consider the simplest system composed of $N=3$ equidistant atoms in a line. Then, we extend our discussion to a larger system composed of $N=5$ atoms. We illustrate our considerations by examining the time evolution of the atomic populations. For the initial conditions we choose the middle (probe) atom to be in its excited state and the other atoms to be in their ground states. We also analyze the directivity function for features indicative of cavity-type modifications of the radiation pattern of the radiation field.

\subsection{Transient regime}\label{sec4a}

\begin{figure}[t]
\includegraphics[width=4.4cm]{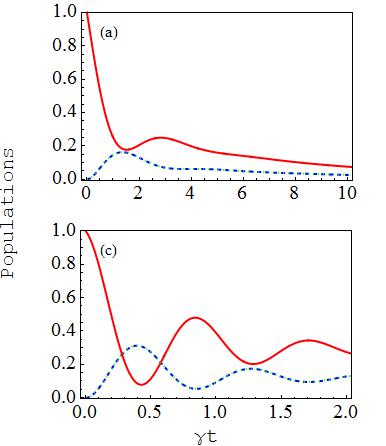}
\includegraphics[width=4.1cm]{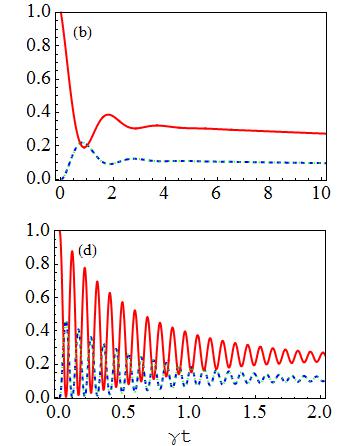}
\caption{(Color online) Time evolution of the atomic populations, $\rho_{22}$ (red solid line), $\rho_{11}$ (blue dashed line), $\rho_{33}$ (green dotted line), plotted for the cavity-type configuration of $3$ atoms and for different distances between the atoms: (a) $r_{12}=r_{23}=\lambda/4$, (b) $r_{12}=r_{23}=\lambda/6$, (c) $r_{12}=r_{23}=\lambda/8$, and (d) $r_{12}=r_{23}=\lambda/16$. Initially, the middle (probe) atom $2$ was excited, $\ket{\psi(0)} =\ket 2$. }
\label{fig8}
\end{figure}

The simplest system which could exhibit features characterizing a cavity formed by atomic mirrors is a chain of $N=3$ equidistant atoms. Let us first examine the process of population transfer between the middle (probe) atom of the chain and the side (mirror) atoms. The time evolution of the populations of the atoms is illustrated in~Fig.~\ref{fig8}, where we present the dependence of the population transfer on the distance between the atoms. It is seen that the population is periodically transferred between the probe atom and the mirror atoms. The transfer occurs with frequency determined by the dipole-dipole coupling $\Omega_{12}\, (=\Omega_{23})$ between the probe and mirror atoms. The oscillations show an interesting behavior that the population transfer is not affected (modulated) by the presence of the dipole-dipole coupling $\Omega_{13}$ between the mirror atoms.

The oscillations are accompanied by a steady decay of the populations. However, depending on the distance between the atoms, the populations may not decay to zero but rather to long-lived non-zero values. It is clearly seen from the figure that for atomic distances $r_{ij}<\lambda/4$, i.e. the spacing between the mirror atoms $r_{13}<\lambda/2$, a significant part of the initial population remains trapped in the probe atom from which it decays very slowly. We have thus a situation similar to that one encounters in a Fabry-P\'erot cavity composed of a pair of parallel plate mirrors~\cite{dowling}. In the cavity, the number of modes for interaction with an atom decreases with an decreasing separation $L$ between the mirrors. For $L<\lambda/2$ the number of modes is suppressed resulting in the suppression of the radiation from the atom.
\begin{figure}[t]
\includegraphics[width=4.4cm]{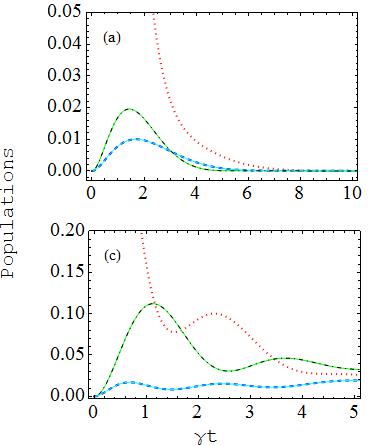}
\includegraphics[width=4.15cm]{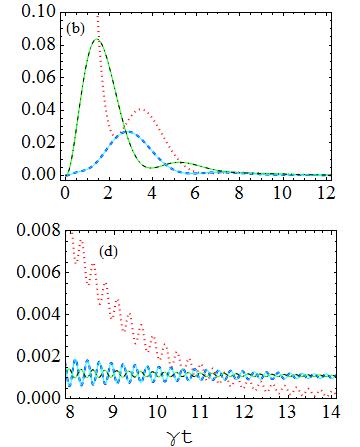}
\caption{(Color online) Time evolution of the atomic populations plotted for the cavity-type configuration of a chain of $5$ atoms and different distances between the atoms: (a) $r_{12}=r_{45}=\lambda/2$, $r_{23}=r_{34}=\lambda$, (b)  $r_{12}=r_{45}=\lambda/4$, $r_{23}=r_{34}=\lambda/2$, (c) $r_{12}=r_{45}=\lambda/6$, $r_{23}=r_{34}=\lambda/3$, and (d) $r_{12}=r_{45}=\lambda/16$, $r_{23}=r_{34}=\lambda/8$. Initially, the middle (probe) atom $3$ was excited, $\ket{\psi(0)} =\ket 3$. Color and style specifications as in Fig.~\ref{fig3}. }
\label{fig9}
\end{figure}

 Figure~\ref{fig9} illustrates the cavity-like situation involving $5$ atoms. Here, each of the cavity mirrors is formed with two atoms. The results are similar to those we encountered for the case of $3$ atoms, Fig.~\ref{fig8}, however, one can see some interesting differences.
 Again, trapping of the population for small distances between the atoms is evident. It is worth noting that the system has a tendency to trap the population in the mirror atoms rather than in the probe atom. It can also be noted that for large distances the probe atom exchanges the population with the next neighbors, the front atoms rather than with the rear atoms forming the mirrors. However, for small distances, the probe atom exchanges the population with the rear atoms leaving the populations of the front atoms almost constant in time. Similar effects appeared to the population transfer in the system composed of a probe atom located at front of an atomic mirror, see Sec.~\ref{sec3a}.

\subsection{Stationary regime}\label{sec4b}

We now turn to the problem of determining the conditions under which the two chains of atoms could act like cavity mirrors to select modes centered about the atomic axis as the only modes available for the emission. We assume that the probe atom is continuously driven by a coherent laser field and consider the field in the steady state limit. 
\begin{figure}[t]
\includegraphics[width=4.26cm]{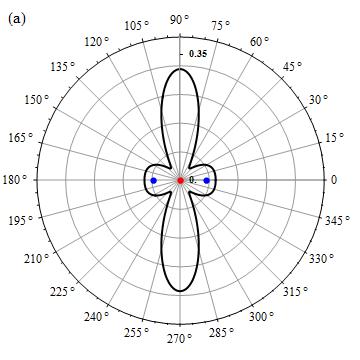}
\includegraphics[width=4.26cm]{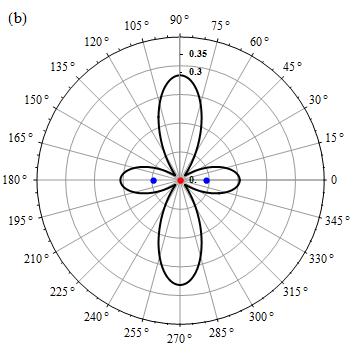}
\includegraphics[width=4.26cm]{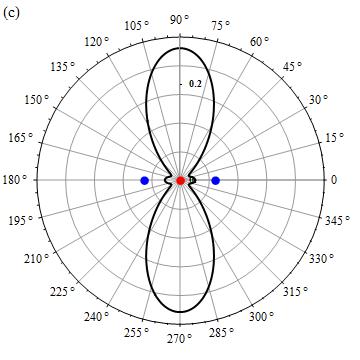}
\includegraphics[width=4.26cm]{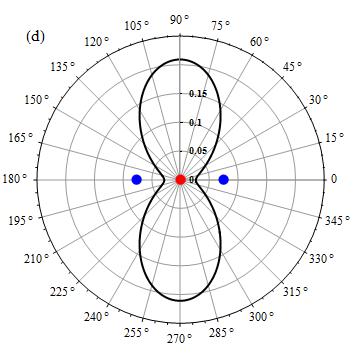}
\caption{(Color online) Polar diagram of the directivity function $D(\theta)$ of the stationary field in the cavity-like situation involving $3$ atoms. The middle atom, which constitutes a probe atom is driven by a cw laser field of the Rabi frequency $\Omega_{0}=0.01\gamma$. The atoms are separated by (a) $r_{ij}=\lambda/2$, (b) $r_{ij}=\lambda/4$, (c) $r_{ij}=\lambda/5$, and (d) $r_{ij}=\lambda/6$.}
\label{fig10}
\end{figure}

Figure~\ref{fig10} illustrates the effect of decreasing distance between atoms on the directivity function $D(\theta)$ of the stationary field for the cavity-like situation involving $3$ atoms. It is seen that the directivity function is very sensitive to the distances between the atoms. For large distances, the system radiates along the cavity axis as well as in the direction normal to the cavity axis. 
When the atomic separations are reduced below $\lambda/4$, which corresponds to the mirror atoms spacing less than $\lambda/2$, we see a cancellation of the radiation along the interatomic axis. Thus, the system turns off the radiation along the cavity axis. It radiates only in the directions normal to the cavity axis, the property which clearly is not characteristic of a cavity.
\begin{figure}[th]

\includegraphics[width=4.22cm]{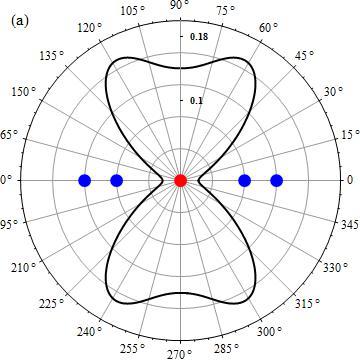}
\includegraphics[width=4.3cm]{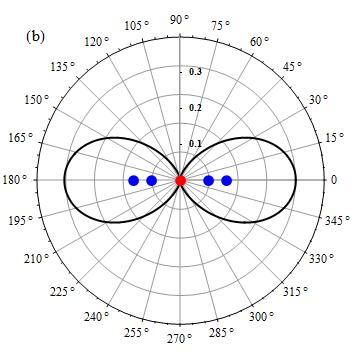}
\includegraphics[width=4.24cm]{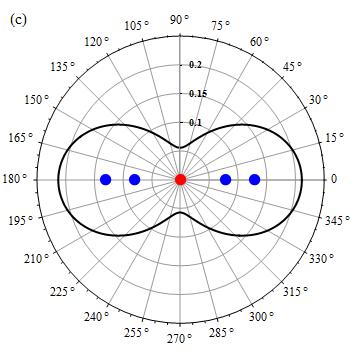}
\includegraphics[width=4.24cm]{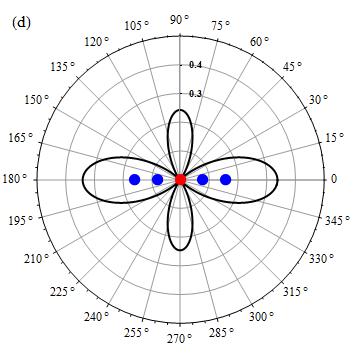}
\caption{(Color online) Polar diagram of the directivity function $D(\theta)$ of the stationary field in the cavity-like situation involving $5$ atoms. The middle atom, which constitutes a probe atom is driven by a cw laser field of the Rabi frequency $\Omega_{0}=0.01\gamma$. In frame (a) the probe atom is separated from its next neighbors by $r_{23}=r_{34}=\lambda/2$, while the atoms which constitute the cavity mirrors are separated by  $r_{12} = r_{45}=\lambda/4$. In frame (b) $r_{23} = r_{34}=\lambda/4$ and $r_{12} = r_{45}=\lambda/8$. In frame (c) $r_{23} = r_{34}=\lambda/4$ and $r_{12} = r_{45}=\lambda/10$. In frame (d) $r_{23} = r_{34}=\lambda/8$ and $r_{12} = r_{45}=\lambda/8$.}
\label{fig11}
\end{figure}

Turning next to the case of $5$ atoms, we plot in Fig.~\ref{fig11} the directivity function for several different distances between the atoms. We see that the directivity function differs significantly from what we observed for the case of three atoms. Thus, the effect of additional atoms in forming the cavity mirrors is clearly more pronounced on the directivity function of the stationary field than on the transient population of the atoms. 
First we note from the figure that there are distances between the atoms at which the directivity function is nonzero only along the interatomic axis [ Figs.~\ref{fig11}(b) and~\ref{fig11}(c) ]. 

It can also be noticed that there is an optimal distance between the mirror atoms at which the solid angle subtended by the cone is minimal, or equivalently the width of the cavity axial mode is very narrow. As seen from Fig.~\ref{fig11}(c), a reduction of the separation may result in an increase of the beamwidth. We also see quite clearly that larger separations between the mirror atoms, additional directions of emission appear  [ Fig.~\ref{fig11}(d) ].

\section{Summary}\label{sec5}

We have studied directional properties of the radiation field emitted by a chain of closely located and dipole-dipole interacting two-level atoms. 
Two geometrical configurations of atoms in the chain have been studied: A probe atom in front of a finite-size chain of closely located atoms, and a probe atom between a pair of chains of closely located atoms. We call the earlier case as an atom in front of an atomic mirror, and the former as an atom inside a cavity whose the mirrors are formed by the two chains of atoms. We have found that it is possible to account for certain 
mirror and cavity-like features, such as 
  We have examined the conditions for one-sided emission centered about the interatomic axis, and have found a lower bound for the distance between the atoms above which a one-sided emission along the interatomic axis can be achieved. The one-sided emission focused into a cone about the interatomic axis and oriented in the backward direction can be regarded as a mirror type with perfect reflectivity. For atomic distances smaller than the lower bound, a part of the population can be trapped in the probe atom indicating that at these distances there are no radiative modes available for the emission. This is a situation similar to that one encounters in a Fabry-P\'erot cavity composed of a pair of parallel plate mirrors~\cite{dowling}.

Polar diagrams have been presented showing the variation of the directivity function of the time dependent as well as the stationary fields with the separation between the atoms.
We have found that the directional properties of the radiation field are generally more manifested in the stationary field when the probe atom is continuously driven by a coherent laser field. 

The control of directionality of photon emission could have applications in quantum information processing in realizing a directional quantum network. For example, the mirror-like behavior of the chain could be used to realize a chiral quantum network, where atoms interact via one-way emission~\cite{pr15,gg15,vr16}. It could also be used as an optical reflector or an optical mirror to create, together with a distant dielectric mirror, a Fabry-P\'erot cavity~\cite{hs11}.

\acknowledgments
Q. G.  acknowledges support from 6th research grant (proposal number 3775) from Jazan University, Saudi Arabia.

\appendix

\section{Atomic correlation functions in the collective state basis}\label{sec6}

The properties of the atomic correlation functions can be studied in terms of the density matrix elements of the operator $\rho$ written in the basis of the free Hamiltonian $H_{0}$ excluding the contribution of the driving field. In the absence of the driving field $(\Omega_{0}=0)$ the eigenstates of the free Hamiltonian $H_{0}$ and their energies are
\begin{align}
\ket 1 &= \ket{g_{1}}\ket{g_{2}}\ket{g_{3}} ,\qquad E_{1} = 0 ,\nonumber\\
\ket 2 &= \ket{g_{1}}\ket{g_{2}}\ket{e_{3}} ,\qquad E_{2} = \hbar\omega_{0} ,\nonumber\\
\ket 3 &= \ket{g_{1}}\ket{e_{2}}\ket{g_{3}} ,\qquad E_{3} = \hbar\omega_{0} ,\nonumber\\
\ket 4 &= \ket{e_{1}}\ket{g_{2}}\ket{g_{3}} ,\qquad E_{4} = \hbar\omega_{0} ,\nonumber\\
\ket 5 &= \ket{g_{1}}\ket{e_{2}}\ket{e_{3}} ,\qquad E_{5} = 2\hbar\omega_{0} ,\nonumber\\
\ket 6 &= \ket{e_{1}}\ket{g_{2}}\ket{e_{3}} ,\qquad E_{6} = 2\hbar\omega_{0} ,\nonumber\\
\ket 7 &= \ket{e_{1}}\ket{e_{2}}\ket{g_{3}} ,\qquad E_{7} = 2\hbar\omega_{0} ,\nonumber\\
\ket 8 &= \ket{e_{1}}\ket{e_{2}}\ket{e_{3}} ,\qquad E_{8} = 3\hbar\omega_{0} .\label{a1}
\end{align}

The natural treatment for interacting atoms is the analysis in the basis of the eigenstates of the Hamiltonian $H_{0}+H_{dd}$, the so-called collective states, which can be expressed in terms of the bare atomic states. The Hamiltonian $H_{0}+H_{dd}\equiv \tilde{H}$ written in the basis of the states (\ref{a1}) has a matrix form
\begin{eqnarray}
\tilde{H} = \hbar\left(
  \begin{array}{cccc}
    0 & 0 & 0 & 0  \\
    0 & \omega_{0}I+M_{1} & 0 & 0 \\
    0 & 0 & 2\omega_{0}I+M_{2} & 0 \\
    0 & 0 & 0 & 3\omega_{0} \\
  \end{array}
\right) ,\label{a2}
\end{eqnarray}
where
\begin{eqnarray}
M_{1} \!=\!\left(\!
  \begin{array}{ccc}
  0 & \Omega_{23} & \Omega_{13} \\
   \Omega_{23} &0 & \Omega_{12}  \\
    \Omega_{13} & \Omega_{12} & 0 \\
  \end{array}\!
\right) ,\nonumber\\
M_{2}\!=\left(\!
  \begin{array}{ccc}
  0 & \Omega_{12} & \Omega_{13} \\
   \Omega_{12} & 0 & \Omega_{23}  \\
   \Omega_{13} & \Omega_{23} & 0 \\
  \end{array}
\right) ,\label{a3}
\end{eqnarray}
and $I$ is the $3\times 3$ unit matrix.
The matrix (\ref{a3}) is block diagonal, composed of two one dimensional blocks and two $3\times 3$ blocks $M_{1}$ and $M_{2}$. 
Thus, the diagonalization of the $8\times 8$ matrix (\ref{a3}) reduces to a diagonalization of the matrices $M_{1}$ and $M_{2}$.
In fact, it is enough to diagonalize the matrix $M_{1}$. The eigenvalues and eigenvectors of the matrix $M_{2}$ are then obtained by replacing $\omega_{0}\rightarrow 2\omega_{0}$ and interchanging $\Omega_{12}\leftrightarrow \Omega_{23}$.

Since $\Omega_{ij}$ is significant for atomic separations $r_{ij}\leq \lambda/4$, we have that at distances $r_{13}> \lambda/2$ we can put $\Omega_{13}=0$ in the matrices $M_{1}$ and $M_{2}$. In this case, a substantial simplification arises in the diagonalization of the matrix $M_{1}$ leading to the following eigenvalues
\begin{align}
\lambda_{2} = \omega_{0} ,\
\lambda_{3} = \omega_{0} + \Omega ,\
\lambda_{4} = \omega_{0} -\Omega ,\label{a4}
\end{align}
and the corresponding eigenstates
\begin{align}
\ket{\tilde{2}} &= -\sin\phi\ket 2 + \cos\phi\ket 4 ,\nonumber\\
\ket{\tilde{3}} &= \frac{1}{\sqrt{2}}\left(\ket s + \ket 3\right) ,\nonumber\\
\ket{\tilde{4}} &= \frac{1}{\sqrt{2}}\left(-\ket s + \ket 3\right) ,\label{a5}
\end{align}
where 
\begin{eqnarray}
\ket{s} = \cos\phi\ket 2 + \sin\phi\ket 4 ,\label{a6}
\end{eqnarray}
with $\sin\phi = \Omega_{12}/\Omega$, $\cos\phi = \Omega_{23}/\Omega$, and $\Omega = \sqrt{\Omega_{12}^{2}+\Omega_{23}^{2}}$.

The eigenvalues of the matrix (\ref{a3}) give the energies of the collective states of the system.

Diagonalize the matrix $M_{1}$. The characteristic equation of the matrix is
\begin{eqnarray}
z^{3} -\left(\Omega_{12}^{2} +\Omega_{23}^{2}\right)z = 0 .\label{q14a}
\end{eqnarray}
where $z=\lambda -\omega_{0}$, and $\lambda$ is an eigenvalue. Equation (\ref{q14a}) can be written as
\begin{eqnarray}
 z\left[z^{2} -\left(\Omega_{12}^{2} +\Omega_{23}^{2}\right)\right] = 0 ,\label{q15a}
\end{eqnarray}
whose the roots (eigenvalues) are
\begin{eqnarray}
z_{1} = 0 ,\ z_{2} = \sqrt{\Omega_{12}^{2}+\Omega_{23}^{2}} ,\
z_{3} = -\sqrt{\Omega_{12}^{2}+\Omega_{23}^{2}} .\label{q16a}
\end{eqnarray}
For the eigenvalue $z=0$, the corresponding eigenstate is
\begin{eqnarray}
\ket{\Psi_{2}} = -\sin\phi\ket 2 + \cos\phi\ket 4 ,\label{q17a}
\end{eqnarray}
where $\sin\phi = \Omega_{12}/\Omega$, $\cos\phi = \Omega_{23}/\Omega$, and $\Omega = \sqrt{\Omega_{12}^{2}+\Omega_{23}^{2}}$.

For the eigenvalues $z=\pm \Omega$, the corresponding eigenstates are
\begin{eqnarray}
\ket{\Psi_{3}} &=& \frac{1}{\sqrt{2}}\left(\ket s + \ket 3\right) ,\qquad z = \Omega ,\nonumber\\
\ket{\Psi_{4}} &=& \frac{1}{\sqrt{2}}\left(-\ket s + \ket 3\right) ,\qquad  z = -\Omega ,\label{q18a}
\end{eqnarray}
where 
\begin{eqnarray}
\ket{s} = \cos\phi\ket 2 + \sin\phi\ket 4 .\label{q19a}
\end{eqnarray}

The eigenvalues and eigenvectors of the matrix $M_{2}$ are obtained by replacing $\omega_{0}\rightarrow 2\omega_{0}$ and interchanging $\Omega_{12}\leftrightarrow \Omega_{23}$ in Eqs.~(\ref{q14a})-(\ref{q18a}). Hence, the eigenstates of the matrix $M_{2}$ and the corresponding energies are 
\begin{align}
\ket{\Psi_{5}} &= -\cos\phi\ket 5 + \sin\phi\ket 7 ,\qquad \lambda_{5} = 2\omega_{0} ,\nonumber\\
\ket{\Psi_{6}} &= \frac{1}{\sqrt{2}}\left(\ket{s^{\prime}} + \ket 3\right) ,\qquad \lambda_{6} = 2\omega_{0} +\Omega ,\nonumber\\
\ket{\Psi_{7}} &= \frac{1}{\sqrt{2}}\left(-\ket{s^{\prime}} + \ket 3\right) ,\qquad  \lambda_{7} = 2\omega_{0} -\Omega ,\label{q18b}
\end{align}
where
\begin{eqnarray}
\ket{s^{\prime}} = \sin\phi\ket 5 + \cos\phi\ket 7 .\label{q19b}
\end{eqnarray}

Having the collective states available, call them $\ket{\Psi_{1}}, \ket{\Psi_{2}},\ldots \ket{\Psi_{8}}$, we can express the atomic correlation functions $\langle S^{+}_{i}S^{-}_{j}\rangle$ in terms of the density matrix elements in the basis of the $\ket{\Psi_{i}}$ states. For example, 
\begin{eqnarray}
 \langle S_{1}^{+}S_{2}^{-}\rangle &=& {\rm Tr}\left\{S_{1}^{+}S_{2}^{-}\rho\right\} \nonumber\\
&=& {\rm Tr}\left\{\sum_{i,j=1}^{8}\rho_{ij}S_{1}^{+}S_{2}^{-}\ket{\Psi_{i}}\bra{\Psi_{j}}\right\} .
\end{eqnarray}
Thus, it requires to calculate the result of $S_{1}^{+}S_{2}^{-}\ket{\Psi_{i}}$ before applying the trace.

The required products of the atomic operators
\begin{align}
&S_{1}^{+}S_{1}^{-} + S_{2}^{+}S_{2}^{-} + S_{3}^{+}S_{3}^{-} = 3\ket 8\!\bra 8 +2\ket 7\!\bra 7 +2\ket 6\!\bra 6 \nonumber\\
&+2\ket 5\!\bra 5 +\ket 4\!\bra 4 +\ket 3\!\bra 3 +\ket 2\!\bra 2 ,\nonumber\\
&S_{1}^{+}S_{2}^{-} + S_{2}^{+}S_{1}^{-} = \ket 6\bra 5 +\ket 5\bra 6 +\ket 4\bra 3 +\ket 3\bra 4 ,\nonumber\\
&S_{1}^{+}S_{2}^{-} - S_{2}^{+}S_{1}^{-} = \ket 6\bra 5 -\ket 5\bra 6 +\ket 4\bra 3 -\ket 3\bra 4 .
\end{align}

We express the bare atomic states in terms of two independent sets of the collective states
\begin{align}
\ket 1 &= \ket{\Psi_{1}} ,\nonumber\\
\ket 2 &= -\sin\phi\ket{\Psi_{2}} +\frac{\cos\phi}{\sqrt{2}}\left(\ket{\Psi_{3}}-\ket{\Psi_{4}}\right) ,\nonumber\\
\ket 3 &= \frac{1}{\sqrt{2}}\left(\ket{\Psi_{3}} +\ket{\Psi_{4}}\right) ,\nonumber\\
\ket 4 &= \cos\phi\ket{\Psi_{2}} +\frac{\sin\phi}{\sqrt{2}}\left(\ket{\Psi_{3}}-\ket{\Psi_{4}}\right) ,
\end{align}
and
\begin{align}
\ket 5 &= -\cos\phi\ket{\Psi_{5}} +\frac{\sin\phi}{\sqrt{2}}\left(\ket{\Psi_{6}}-\ket{\Psi_{7}}\right) ,\nonumber\\
\ket 6 &=  \frac{1}{\sqrt{2}}\left(\ket{\Psi_{6}} +\ket{\Psi_{7}}\right) ,\nonumber\\
\ket 7 &= \sin\phi\ket{\Psi_{5}} +\frac{\cos\phi}{\sqrt{2}}\left(\ket{\Psi_{6}}-\ket{\Psi_{7}}\right) ,\nonumber\\
\ket 8 &= \ket{\Psi_{8}} .
\end{align}

Hence
\begin{align}
&S_{1}^{+}S_{1}^{-} + S_{2}^{+}S_{2}^{-} = P_{22}\cos^{2}\phi + \frac{1}{2}\!\left(1+\sin^{2}\phi\right)\!\left(P_{33} +P_{44}\right) \nonumber\\
&+{\rm Re}\!\left[P_{34}\cos^{2}\phi +\frac{\sin2\phi}{\sqrt{2}}\left(P_{23} -P_{24}\right)\right] ,\nonumber\\
&S_{1}^{+}S_{2}^{-} + S_{2}^{+}S_{1}^{-} = \left(P_{33}-P_{44}+P_{66}-P_{77}\right)\sin\phi \nonumber\\
&+\sqrt{2}\,{\rm Re}\!\left[P_{23}+P_{24}-P_{56}-P_{57}\right]\cos\phi ,\nonumber\\
&S_{1}^{+}S_{2}^{-} - S_{2}^{+}S_{1}^{-} = 2\,{\rm Im}\!\left[P_{43}+P_{76}\right]\sin\phi \nonumber\\
&+\sqrt{2}\,{\rm Im}\!\left[P_{56}+P_{57}+P_{32}+P_{42}\right]\cos\phi ,\label{A18}
\end{align}
where $P_{ij}=\ket{\Psi_{i}}\bra{\Psi_{j}}$. 

It is clearly seen from Eq.~(\ref{A18}) that the real parts of the atomic correlations functions depend on the populations of the collective states and coherences between them, while the imaginary parts depend solely on the coherences. Thus, crucial for the imaginary parts of the atomic correlation functions to be nonzero is to prepare or drive the atomic system such that there are nonzero coherences between the collective states.

\section{Equations of motion}\label{sec6a}
Assuming only one atom at the most can be excited at any time $t$, the time-dependent state vector can be written as

\begin{equation}
\label{state-vec}
 |\psi (t)\rangle= \sum^N_{i=1} b_i(t)|i\rangle+  b_{N+1}(t)|N+1\rangle\,,
\end{equation}

where the ket vector $ |i \rangle$ represents the combined state of the $N$ atoms, respectively with only the $i$th atom excited. $ |N+1\rangle$ represents the collective ground state of all $N$ atoms with no atom excited.

The master equation~(\ref{q1}) readily renders the equations of motion for the populations, i.e., the diagonal density matrix elements~(\ref{population}a-\ref{population}b) as well as the coherences~(\ref{population}c-\ref{population}d).

\begin{subequations}
\label{population}
\begin{align}
&\dot{ \rho}_{l,l}(t)=\frac{d}{dt} |b_l(t) |^2=-\gamma  |b_l(t) |^2-\sum_{
 j\neq l  }\nonumber\\
&\bigg[\bigg(i \Omega^{(lj)}+\frac{\gamma^{(lj)}}{2}\bigg)b_j^*(t)b_l(t)+h.c.\bigg]\,,\\
&\dot{ \rho}_{N+1,N+1}(t)=\frac{d}{dt} |b_{N+1}(t) |^2 =\frac{1}{2}\sum^N_{ i=1 }\sum^N_{ j=i+1 }
\bigg[\big(\gamma^{(ij)}+\gamma^{(ji)}\big)\nonumber\\&\big(b_i^*(t)b_j(t)+h.c.\big)\bigg]+\gamma\sum^N_{j=1} |b_j(t) |^2\,,\\
&\dot{ \rho}_{m,n}(t)\bigg|_{\substack{\{m,n\}\in\{1,\dots,N\} \\ m\neq n}}=\frac{d}{dt}( b_m^*(t)b_n(t))=-\gamma b_m^*(t)b_n(t)\nonumber\\&-\sum^N_{\substack{
  j=1 \\
  j\neq m
  }}
\bigg(i \Omega^{(mj)}+\frac{\gamma^{(mj)}}{2}\bigg)b_j^*(t)b_n(t)\nonumber\\&+\sum^N_{\substack{
  j=1 \\
  j\neq n
  }}
\bigg(i \Omega^{(nj)}-\frac{\gamma^{(nj)}}{2}\bigg)b_m^*(t)b_j(t)\,,\\
&\dot{ \rho}_{l,N+1}(t)=-\sum^N_{\substack{
   j=1 \\
  j\neq l
  }}\bigg(i \Omega^{(lj)}+\frac{\gamma^{(lj)}}{2}\bigg)b_j^*(t)b_{N+1}(t)\nonumber\\&-\frac{\gamma}{2}b_l^*(t)b_{N+1}(t),
\end{align}
\end{subequations}

where $\{l,m,n\}\in\{1,\dots,N\}$. The remaining equations can be obtained from~(\ref{population}c-\ref{population}d)  by complex conjugation.

\end{document}